\documentstyle[sprocl]{article}
\begin{document}
\begin{flushright}{UT-747\ \ ,\ \ 1996}\end{flushright}
\vskip 1.5 truecm
\title{PHASE OPERATOR PROBLEM AND AN INDEX THEOREM FOR Q-DEFORMED OSCILLATOR\footnote{To be published in the Proceedings of International Symposium on 
Frontiers in Quantum Field Theory in Honor of Keiji Kikkawa, Toyonaka, Osaka,
December 14-17, 1995 (World Scientific, Singapore)}}
\author{Kazuo Fujikawa}
\address{Department of Physics, University of Tokyo\\
Bunkyo-ku, Tokyo 113, Japan}
\maketitle\abstracts{
  The notion of  index is applied to analyze the phase operator problem 
associated with the photon.  We clarify the absence of the 
hermitian phase operator on the basis of an index consideration. 
We point out an interesting analogy between the phase operator problem and the 
chiral anomaly in gauge theory, and an appearance of a new class of quantum anomaly is noted.
 The notion of index, which is  invariant under a wide class of continuous 
deformation,  is also shown to be useful to  characterize the representations 
of Q-deformed oscillator algebra. }

\section{Phase operator}
\par
The notion of the phase operator was introduced by Dirac in his study of the 
second quantization of the photon field[1]. The free classical Maxwell 
field is equivalent to an  infinite set of oscillators
\begin{equation}
H = \frac{1}{2}\sum_{\vec{k},\lambda} \omega (\vec{k})[a^{\star}_{\vec{k}\lambda}a_{\vec{k}\lambda} + a_{\vec{k}\lambda}a^{\star}_{\vec{k}\lambda}] 
\end{equation}
and the essence of the phase operator is understood by considering 
 the simplest one-dimensional harmonic oscillator 
\begin{eqnarray}
h&=&\frac{1}{2} (a^{\dagger}a + aa^{\dagger}) \nonumber\\
 &=&  a^{\dagger}a + 1/2
\end{eqnarray}
where $a$ and $a^{\dagger}$ stand for the annihilation and creation
operators satisfying  the standard commutator
\begin{equation}
[a, a^{\dagger} ] = 1
\end{equation}
The vacuum state $|0\rangle$
is annihilated by $a$
\begin{equation}
a|0 \rangle = 0
\end {equation}
which ensures the absence of  states with negative norm.
The number operator defined by
\begin{equation}
N = a^{\dagger}a
\end{equation}
then has non-negative integers as eigenvalues, and the annihilation
operator
$a$ is represented by
\begin{equation}
a  = |0\rangle \langle 1|+ |1\rangle \langle 2|\sqrt{2}
              + |2\rangle \langle 3|\sqrt{3} + ....
\end{equation}
in terms of the eigenstates $|k\rangle$ of the number operator
\begin{equation}
N|k\rangle = k|k\rangle
\end{equation}
with $k = 0, 1, 2, ... $. The creation operator $a^{\dagger}$ is given
by the hermitian conjugate of $a$ in (6).\\

\noindent
{\em Index and Phase Operator}\\

In the representation of $a$ and $a^{\dagger}$ specified above we have the
index condition
\begin{equation}
dim\ ker\ a^{\dagger}a - dim\ ker\ aa^{\dagger} = 1
\end{equation}
where $dim\ ker\ a^{\dagger}a$ ,for example, stands for the number of
normalizable basis vectors $u_{n}$ which satisfy $a^{\dagger}au_{n}=0$
;$dim\ ker\ a^{\dagger}a$ thus agrees with the number of zero eigenvalues 
of the hermitian operator $a^{\dagger}a$.
In the conventional notation of index theory, the relation (8) is 
written by using the trace of well-defined operators as
\begin{equation}
Tr(e^{-a^{\dagger}a/M^{2}}) - Tr(e^{-aa^{\dagger}/M^{2}}) = 1
\end{equation}
with $M^{2}$ standing for a positive constant. 
 The relation(9) is confirmed for the standard representation(6) as
\begin{equation}
1 + (\sum_{n=1}^{\infty}e^{-n/M^{2}}) - (\sum_{n=1}^{\infty}e^{-n/M^{2}})= 1
\end{equation}
independently of the value of $M^{2}$. 

If one should suppose the existence of a well defined hermitian phase operator
$\phi$, one would have a polar decomposition 
\begin{equation}
a = U(\phi)H = e^{i\phi}H
\end{equation}
as was originally suggested by Dirac[1]. Here  $U$ and $H$ 
stand for unitary and hermitian operators, respectively.
If (11) should be valid, one has
\begin{equation}
aa^{\dagger} = UH^{2}U^{\dagger}
\end{equation}
which is unitary equivalent to $a^{\dagger}a = H^{2}$; 
$a^{\dagger}a$ and $aa^{\dagger}$ thus have an
identical number of zero eigenvalues. In this case, we have in the same 
notation as (9)
\begin{eqnarray}
&&Tr(e^{-a^{\dagger}a/M^{2}}) - Tr(e^{-aa^{\dagger}/M^{2}})\nonumber\\
&=&Tr(e^{-H^{2}/M^{2}}) - Tr(e^{-UH^{2}U^{\dagger}/M^{2}})= 0
\end{eqnarray}
This relation when combined with (9) constitutes a proof of the absence of 
a hermitian phase operator in the framework of index theory[2].

In the classical level, one may define 
\begin{eqnarray}
a &=& e^{i\phi}\sqrt{N}\nonumber\\
  &=& \frac{1}{\sqrt{2}}(ip + q)
\end{eqnarray}
or
\begin{eqnarray}
p &=& \sqrt{N}\sin\phi\nonumber\\
q &=& \sqrt{N}\cos\phi
\end{eqnarray}
This transformation is confirmed to be canonical, 
and thus one has a Poisson bracket
\begin{equation}
\{ N, \phi\}_{PB} = 1
\end{equation}
which naively leads to a commutator 
\begin{equation}
[\hat{N}, \hat{\phi}] = i
\end{equation}
and the well-known uncertainty relation
\begin{equation}
\Delta N\Delta \phi \geq \frac{1}{2}
\end{equation}
The absence of the hermitian phase operator $\phi$ shows that this uncertainty 
relation has no mathematical basis[3].  

The basic utility of the notion of index or an index theorem lies
in the fact that the index as such is an integer and remains invariant
under a wide class of continuous deformation. 
For  example, the unitary time development of $a$ and $a^{\dagger}$ dictated
by the Heisenberg equation of motion, which includes a fundamental phenomenon
such as squeezing, does not alter the index relation.
Another consequence is that  one cannot generally  relate the representation 
spaces of annihilation operators with {\em different} indices by a 
unitary transformation.

\section{Various phase operators and their implications}

Although the hermitian phase operator , as was introduced by Dirac, does not 
exist, there exist various interesting proposals of alternative  
operators which may replace the phase operator in practical applications. We 
here commnet on two representative ones from the view point of index relation.\\

\noindent
{\em 2.1  Phase Operator of Susskind-Glogower}\\

The phase operator of Susskind and Glogower[3] is defined by
\begin{eqnarray}
e^{i\varphi}&=& \frac{1}{\sqrt{N+1}}a\nonumber\\
&=& |0\rangle \langle 1|+ |1\rangle \langle 2|
              + |2\rangle \langle 3| + ....
\end{eqnarray}
in terms of the eigenstates $|k\rangle$ of the number operator in (7).
This phase operator is related to the operator $a$ in (6) by 
$a=e^{i\varphi}N^{1/2}$. The above operator satisfies the relations
\begin{eqnarray}
e^{i\varphi}(e^{i\varphi})^{\dagger}& =& 1\nonumber\\
(e^{i\varphi})^{\dagger}e^{i\varphi}& =& 1 - |0\rangle\langle 0| \neq 1
\end{eqnarray}
and apparently $\varphi$ is not hermitian.

The analogues of cosine and sine functions, which are hermitian and thus observable,  are defined by
\begin{eqnarray}
C(\varphi)&=& \frac{1}{2}(e^{i\varphi} + (e^{i\varphi})^{\dagger})\nonumber\\
S(\varphi)&=& \frac{1}{2i}(e^{i\varphi} - (e^{i\varphi})^{\dagger})
\end{eqnarray}
These cosine and sine operators satisfy an anomalous commutator 
\begin{equation}
{[}C(\varphi) , S(\varphi)] =  \frac{1}{2i}|0\rangle \langle 0|
\end{equation}
and an anomalous identity
\begin{equation}
C(\varphi)^{2} + S(\varphi)^{2} = 1 - \frac{1}{2}|0\rangle \langle 0|
\end{equation}
The modified trigonometric
operators also satisfy the commutation relations with the number operator $N$,
\begin{eqnarray}
{[} N  , C(\varphi)]        & = & -i S(\varphi) \nonumber \\
{[} N  , S(\varphi)]        & = &  i C(\varphi) 
\end{eqnarray}

For a representation with non-negative $N$ in (19),one 
obtains
the index relation
\begin{eqnarray}
dim\ ker\ e^{i\varphi} - dim\ ker\ (e^{i\varphi})^{\dagger}
=dim\ ker\ a - dim\ ker\ a^{\dagger}=1
\end{eqnarray}
namely, the operator $e^{i\varphi}$ carries a unit index. The index relation
written in the form  
\begin{equation}
dim\ ker\ (e^{i\varphi})^{\dagger}e^{i\varphi} - dim\ ker\ e^{i\varphi}
(e^{i\varphi})^{\dagger} = 1
\end{equation}
which is in agreement with  (20),  is directly related to the above anomalous commutator 
(22) and the anomalous identity(23).\\

\noindent
{\em 2.2 Phase Operator of Pegg and Barnett}\\

One may consider an $s+1$ dimensional truncation of annihilation and 
creation operators
\begin{eqnarray}
a_{s} &=&|0\rangle\langle 1| + |1\rangle\langle 2|\sqrt{2}
                         + .... + |s-1\rangle\langle s|\sqrt{s}\nonumber\\
a_{s}^{\dagger} &=&|1\rangle\langle 0| + |2\rangle\langle 1|\sqrt{2}
                         + .... + |s\rangle\langle s-1|\sqrt{s}
\end{eqnarray}
and then let $s\rightarrow large$ later. In this case, one has the index
condition
\begin{eqnarray}
Tr_{s+1}(e^{-a_{s}^{\dagger}a_{s}/M^{2}}) - 
Tr_{s+1}(e^{-a_{s}a_{s}^{\dagger}/M^{2}})= 0
\end{eqnarray}
in terms of $s+1$ dimensional trace $Tr_{s+1}$; $ker \ a^{\dagger}a = \{ 
|0\rangle \}$ and $ ker \ aa^{\dagger} = \{ |s\rangle \}$. In fact one can confirm that any finite dimensional square matrix carries a 
vanishing index. The operators $a_{s}^{\dagger}$ and $a_{s}$ thus satisfy a 
necessary condition for the existence of a hermitian phase operator. 
Pegg and Barnett proposed[4] to define a hermitian phase operator by
\begin{equation}
e^{i\phi}=|0\rangle \langle 1| + |1\rangle \langle 2| + ...
          + |s-1\rangle \langle s|
          + e^{i(s+1)\phi_{0}}|s\rangle \langle 0|
\end{equation}
where $\phi_{0}$ is an arbitrary constant c-number. The operator $e^{i\phi}$,
which satisfies $a_{s} = e^{i\phi}\sqrt{N}$, 
is in fact unitary in  $s+1$ dimensions
\begin{equation}
e^{i\phi}(e^{i\phi})^{\dagger}=(e^{i\phi})^{\dagger}e^{i\phi}=1
\end{equation}
One may then define cosine and sine operators by
\begin{eqnarray}
\cos\phi &=& \frac{1}{2} (e^{i\phi} + e^{-i\phi})\nonumber\\
\sin\phi &=& \frac{1}{2i}(e^{i\phi} - e^{-i\phi})
\end{eqnarray}
with $e^{-i\phi} = (e^{i\phi})^{\dagger}$.
These operators together with the number operator satisfy the commutation
relations
\begin{eqnarray}
{[} N, \cos\phi] &=& -i\sin\phi\nonumber\\
&+&\frac{(s+1)}{2}[e^{i(s+1)\phi_{0}}|s\rangle \langle 0|
                 -e^{-i(s+1)\phi_{0}}|0\rangle \langle s|]\nonumber\\
{[} N, \sin\phi] &=&  i\cos\phi\nonumber\\
&-&i\frac{(s+1)}{2}[e^{i(s+1)\phi_{0}}|s\rangle \langle 0|
                 +e^{-i(s+1)\phi_{0}}|0\rangle \langle s|]
\end{eqnarray}
and
\begin{eqnarray}
{[} \cos\phi, \sin\phi] &=& 0\nonumber\\
\cos^{2}\phi + \sin^{2}\phi &=& 1
\end{eqnarray}
The last expression shows that these operators satisfy normal algebraic
relations.\\

\noindent
{\em 2.3 Physical Implications of Various Phase Operators}\\

The state $|s\rangle$ is responsible for the vanishing index in (28) and 
thus for the existence of the hermitian phase operator in (29). In other words,
the hermiticity critically depends on the state $|s\rangle$ and the phase
operator is cut-off sensitive. The state $|s\rangle$ does not decouple even 
for large cut-off parameter $s$ from the hermitian phase operator $\phi$.
It has been  shown[2] that the algebraic consistency, i.e., the presence or  absence   
of minimum uncertainty states is a good test of the effects of $|s\rangle$ and
the index idea.

In fact, one can show the uncertainty relations for a physical state 
$|p\rangle$
\begin{eqnarray}
\Delta N \Delta \sin\phi \geq \Delta N \Delta S(\varphi)
             \geq \frac{1}{2}|\langle p|C(\varphi)|p\rangle|
             =   \frac{1}{2}|\langle p|\cos\phi|p\rangle|\\
\Delta \cos\phi \Delta \sin\phi \geq \Delta C(\varphi) \Delta S(\varphi)
         \geq \frac{1}{4}|\langle p|0\rangle \langle 0|p\rangle \geq 0
\end{eqnarray}
These relations show that the uncertainty relations are always better satisfied for the operators  $C(\varphi)$ and $S(\varphi)$ of 
Susskind and Glogower which exhibits ``anomalous `` relations (22) and (23). In particular,for the states with small average photon 
numbers, the relation in (34), for example, becomes an inequality[2]
\begin{eqnarray}
\Delta N \Delta \sin\phi > \Delta N \Delta S(\varphi)
             \geq \frac{1}{2}|\langle p|C(\varphi)|p\rangle|
             =   \frac{1}{2}|\langle p|\cos\phi|p\rangle|
\end{eqnarray}
due to the effect of the state $|s\rangle$ even for arbitrarily large $s$.
In other words, one cannot generally judge if the measurement is  at the 
quantum limit or not if one  uses the phase operator $\phi$ of Pegg and Barnett, since one cannot achieve the minimum uncertainty in terms of $\phi$.
The operator $\varphi$, which satisfies anomalous relations in (22) and (23)
, is in fact more consistent quantum mechanically; one can judge the quantum
limit of measurements by looking at the uncertainty relation (36).

\section{Analogy with chiral anomaly and a new class of quantum anomaly}

It is known[5] that the chiral anomaly is characterized by  the 
Atiyah-Singer index theorem[6]
\begin{equation}
Tr e^{-a^{\dagger}a} - Tr e^{-aa^{\dagger}} = \nu
\end{equation}
where the Pontryagin index $\nu$ is expressed as an integral of 
$ F\tilde{F}$, if one defines $a$ by the chiral Dirac operator 
\begin{equation}
a = \gamma^{\mu}{\cal D}_{\mu}(\frac{1+\gamma_{5}}{2})
\end{equation}
A failure of a unitary transformation to an interaction picture specified by
a vanishing index 
\begin{equation}
Tr e^{-a_{0}^{\dagger}a_{0}} - Tr e^{-a_{0}a_{0}^{\dagger}} = 0
\end{equation}
with $a_{0} = \gamma^{\mu}\partial_{\mu}(1+\gamma_{5})/2$ leads to the anomalous
behavior of perturbation theory. Also, the failure of decoupling of the cut-off parameter such as the Pauli-Villars regularization mass is
well known: If the regulator mass is kept finite one obtains a normal relation,but in the limit of infinite regulator mass, the mass term gives rise to the 
chiral anomaly.

The failure of the decoupling of cut-off also takes place in the phase operator . One may rewrtie the index relation (28) as
\begin{equation}
Tr_{s+1} e^{-a_{s}^{\dagger}a_{s}/M^{2}} - Tr_{s} e^{-a_{s}a_{s}^{\dagger}/M^{2}} =
Tr_{s+1} e^{-a_{s}a_{s}^{\dagger}} - Tr_{s} e^{-a_{s}a_{s}^{\dagger}} = 1
\end{equation}
where $Tr_{s}$ stands for a trace over the first $s$ dimensional subspace, and
the right hand side of this equation comes from the contribution of 
the state $|s\rangle$. By taking $s \rightarrow \infty$ in this expression, one
recovers the index condition (9).
The effect of the cut-off parameter $|s\rangle$ in the limit $s \rightarrow \infty$ gives 
rise to the non-vanishing index; this is the same phenomenon as the chiral 
anomaly in the Pauli-Villars regularization.

It has been  proposed  to regard the non-vanishing index (9) and the associated 
anomalous behavior in (22) and (23) as a new class of quantum anomaly[2], which arises from
an infinite number of degrees of freedom associated with the Bose statictics;
one can put an arbitrary number of photons in a specified quantum state. 
In this view point, the absence of the hermitian phase operator is an 
inevitable quantum effect, {\em not} an artifact of our insufficient definition
of the phase operator. We have shown in (36) that the apparently anomalous 
behavior exhibited by the operator $\varphi$ in (22) and (23) is in fact more
consistent with the principle of quantum mechanics.

\section{Index as an invariant characterization of Q-deformed oscillator algebra}

The index such as 
\begin{equation}
dim\ ker\ a - dim\ ker\ a^{\dagger} = 1
\end{equation}
is expected to be invariant under a wide class of continuous deformation 
of the operator $a$. The index may thus provide an invariant characterization of representations of Q-deformed oscillator algebra[7]. We analyze  this problem by
considering the Q-deformed algebra
\begin{eqnarray}
[a, a^{\dagger}] &=& [N+1] - [N]\nonumber\\
{[N, a^{\dagger}]} &=& a^{\dagger}\nonumber\\
{[N, a]} &=& - a
\end{eqnarray}
where
\begin{equation}
[N] \equiv \frac{q^{N} - q^{-N}}{q - q^{-1}}
\end{equation}
with $q$ standing for a deformation parameter. The above algebra satisfies 
the Hopf structure, and the  Casimir operator for the algebra (42) is 
given by[8] 
\begin{equation}
C = a^{\dagger}a - [N]
\end{equation}
and thus $[N]$ and $a^{\dagger}a$ are independent in general. Practically the
above Q-deformed oscillator  is useful to study a Q-deformed $SU(2)$ in the Schwinger construction[9].

To deal with a general situation, we start with a matrix representation
of the above algebra defined by
\begin{eqnarray}
a &=& \sum_{k=1}^{\infty} \sqrt{[k]}|k-1\rangle\langle k|\nonumber\\
a^{\dagger} &=& \sum_{k=1}^{\infty} \sqrt{[k]}|k\rangle\langle k-1|\nonumber\\
N &=& \sum_{k=0}^{\infty} k|k\rangle\langle k|\nonumber\\
C &=& a^{\dagger}a - [N] = 0
\end{eqnarray}
where the state vectors stand for column vectors such as
\begin{equation}
 |0\rangle =\left(
 \begin{array}{c}
  1\\ 0\\ 0\\ 0\\.
 \end{array}
 \right), \ \
 |1\rangle =\left(
 \begin{array}{c}
  0\\ 1\\ 0\\.\\.
 \end{array}
 \right),
\end{equation}
We examine several cases specified by a different deformatiom parameter $q$.\\

\noindent
I. $q > 0$\\

In this case, $[k] \neq 0$ for $ k \geq 1$, and one has an index condition 
\begin{equation}
dim\ ker\ a - dim\ ker\ a^{\dagger} = 1
\end{equation}
since $ker\ a = \{ |0\rangle \}$ and $ker\ a^{\dagger} = empty$. We thus have no hermitian phase operator, and the operator defined by
\begin{eqnarray}
e^{i\varphi} &=& \frac{1}{\sqrt{[N+1]}}a\nonumber\\
&=& |0\rangle\langle 1| + |1\rangle\langle 2| + ...
\end{eqnarray}
becomes identical to that of Susskind and Glogower. Namely, not only the 
index but also the phase operator itself are invariant under Q-deformation[10].

The deformation with $|q| = 1$ is also known to be allowed. We thus examine\\

\noindent
II-1, $q = e^{2\pi i\theta}, \theta = irrational$\\

In this case, 
\begin{equation}
[k] = \frac{q^{k} - q{-k}}{q - q^{-1}} = \frac{\sin 2\pi k\theta}
{\sin 2\pi\theta} \neq 0
\end{equation}
for $k \geq 1$, and thus the index as well as the phase operator are the 
same as for $q > 0$.\\

\noindent
II-2, $q = e^{2\pi i\theta}, \theta = rational$\\

In particular, we concentrate on the case where $\theta$ is a primitive root of
unity[11],
\begin{equation}
\theta = \frac{1}{s+1}
\end{equation}
with $s$ standing for a natural number. In this case, we have
\begin{equation}
[s+1] = \frac{q^{s+1} - q^{-(s+1)}}{q - q^{-1}} = 0
\end{equation}
and the representation in (45) becomes
\begin{eqnarray}
a &=& \sqrt{[1]}|0\rangle\langle 1| + ... + \sqrt{[s]}|s-1\rangle\langle s|\nonumber\\
&+& \sqrt{[1]}|s+1\rangle\langle s+2| + ... + \sqrt{[s]}|2s\rangle\langle 2s+1|\nonumber\\
&+& .....\nonumber\\
a^{\dagger} &=& \sqrt{[1]}|1\rangle\langle 0| + ... + \sqrt{[s]}|s\rangle\langle s-1|\nonumber\\
&+& \sqrt{[1]}|s+2\rangle\langle s+1| + ... + \sqrt{[s]}|2s+1\rangle\langle 2s|\nonumber\\
&+& .....\nonumber\\
N &=& |1\rangle\langle 1| + 2|2\rangle\langle 2| + ......+s|s\rangle\langle s|\nonumber\\
&+& (s+1)|s+1\rangle\langle s+1| + ..... +(2s+1)|2s+1\rangle\langle 2s+1|\nonumber\\
&+& .....
\end{eqnarray}
Note that 
\begin{equation}
a^{\dagger} \neq (a)^{\dagger}
\end{equation}
which means that the norm of the representation space is not positive definite.

In the representation (52),we have
\begin{eqnarray}
ker \ a &=& \{ |0\rangle, |s+1\rangle, .....\}\nonumber\\
ker \ a^{\dagger} &=& \{ |s\rangle, |2s+1\rangle, ....\}
\end{eqnarray}
and thus 
\begin{equation}
dim\ ker\ a = \infty, \ \ dim\ ker\ a^{\dagger} = \infty
\end{equation}
>From a view point of index , this situation is singular. One may regard this as an indication that the deformation is discontinuous, and thus the notion of 
index looses its precise meaning. In fact, one may understand the above 
situation as 
\begin{equation}
dim\ ker\ a - dim\ ker\ a^{\dagger} = 1
\end{equation}
with  the non-hermitian phase operator 
\begin{equation}
e^{i\varphi} = |0\rangle\langle 1| + |1\rangle\langle 2| + ....
\end{equation}
or as
\begin{equation}
dim\ ker\ a - dim\ ker\ a^{\dagger} = 0
\end{equation}
with a hermitian phase operator[7]
\begin{eqnarray}
e^{i\Phi} &=& |0\rangle\langle 1| + .... +e^{i\phi_{0}}|s\rangle\langle 0|\nonumber\\
&+& |s+1\rangle\langle s+2| + ... + e^{i\phi_{1}}|2s+1\rangle\langle s+1|\nonumber\\
&+& .......
\end{eqnarray}
Both of these operators satisfy
\begin{eqnarray}
a &=& e^{i\varphi}\sqrt{[N]}\nonumber\\
  &=& e^{\Phi}\sqrt{[N]}
\end{eqnarray}
\\
\noindent
{\em  More Conventional Interpretation}\\

A more conventional interpretation of the case $q = \exp [2\pi i/(s+1)]$ is to consider a basic $s+1$ dimensional Weyl block 
\begin{eqnarray}
a_{s} &=& \sqrt{[1]}|0\rangle\langle 1| + ..... +\sqrt{[s]}|s-1\rangle\langle s|,\nonumber\\
a_{s}^{\dagger} &=& \sqrt{[1]}|1\rangle\langle 0| + ..... +\sqrt{[s]}|s\rangle\langle s-1|,\nonumber\\
N_{s} &=& |1\rangle\langle 1| + .......+ s|s\rangle\langle s|,\nonumber\\
C &=& 0 
\end{eqnarray}
with the index
\begin{equation}
dim\ ker\ a_{s} - dim\ ker\ a_{s}^{\dagger} = 0
\end{equation}
One may adopt the hermitian phase operator of Pegg and Barnett[4] for this case. 
However, one cannot recover a phase operator for the conventional case with 
$q=1$ by taking the limit $s\rightarrow \infty$ of the above case: First of 
all, we have a negative norm, as is indicated by $a_{s}^{\dagger} \neq 
(a_{s})^{\dagger}$, and also the phase operator is index sensitive and the 
limit $s\rightarrow \infty$ is ill-defined.

It is interesting to note that one can define a representation without a negative norm for $q = \exp [2\pi i/(s+1)]$ by using the freedom associated with the Casimir operator. One may choose the Casimir operator in (42) as[7]
\begin{equation}
C = [n_{0}] = \frac{1}{\sin (\frac{2\pi}{s+1})}
\end{equation}
with $n_{0} = (s+1)/4$. One may then define the representation of (42) with a positive 
definite norm by
\begin{eqnarray}
a_{s} &=& \sqrt{[1-n_{0}] + [n_{0}]}|0\rangle\langle 1| + .....
        + \sqrt{[s-n_{0}] + [n_{0}]}|s-1\rangle\langle s|\nonumber\\
a_{s}^{\dagger} &=& (a_{s})^{\dagger}\nonumber\\
N_{s} &=& -n_{0}|0\rangle\langle 0| + .... + (s-n_{0})|s\rangle\langle s|
\end{eqnarray}
Note that the arguments inside the square root in (64) are all non-negative.
The index is vanishing in this case, and one may adopt  the hermitian 
phase operator of Pegg and Barnett in (29) which satisfies
\begin{equation}
a = e^{i\phi}\sqrt{[N] + [n_{0}]}
\end{equation}
It is however clear in this case that one cannot recover the normal case with
$q=1$ and $C=0$ by taking the limit $s\rightarrow \infty$. The deformation away from $q=1$ is apparently discontinuous.

\section{Conclusion}
We may conclude our analyses of the phase operator on the basis of the notion of index as follows:\\

\noindent
(i)The notion of index is useful for the analysis of the phase operator and also for an invariant characterization of representations of the Q-deformed 
algebra, if 
\begin{eqnarray}
dim\ ker\ a < \infty\nonumber\\
dim\ ker\ a^{\dagger} < \infty
\end{eqnarray}
This corresponds to the case with a real and positive deformation parameter $q$. We also encounter a new calss of quantum anomaly specified by a non-trivial 
index, which is associated with the Bose statistics.\\

\noindent
(ii)For $q = e^{2\pi i \theta}$ with a real $\theta$, in particular for a rational $\theta$, the index  becomes 
singular . 
\begin{eqnarray}
dim\ ker\ a = \infty\nonumber\\
dim\ ker\ a^{\dagger} = \infty
\end{eqnarray}
The singular index is still useful since it indicates that the deformation
\begin{equation}
q = 1 \rightarrow q = e^{2\pi i \theta}
\end{equation}
is a {\em discontinuous} deformation.\\

\noindent
(iii)The index is modified by a change of the asymptotic behavior of the 
potential. For example, one may consider(here $q$ stands for a coordinate)
\begin{eqnarray}
a &=& \frac{1}{\sqrt{2}}( \frac{\partial}{\partial q} + q + 3cq^{2})\nonumber\\
a^{\dagger} &=& \frac{1}{\sqrt{2}}( - \frac{\partial}{\partial q} + q + 3cq^{2})
\end{eqnarray}
with a positive constant $c$. The conventional harmonic oscillator corresponds to $c=0$. In the case (69) one can show that
\begin{eqnarray}
dim\ ker\ a - dim\ ker\ a^{\dagger} &=& 0 \nonumber\\
dim\ ker\ a = dim\ ker\ a^{\dagger} &=& 0
\end{eqnarray}
and thus one can define a well-defined hermitian phase operator[12]. This is 
somewhat analogous to the Witten index to characterize supersymmetry; one 
can generally change the index if one modifies the asymptotic behavior of the
potential.

\end{document}